\documentclass[english,aps,prb,twocolumn,nofootinbib,showpacs,superscriptaddress,amsmath,amssymb,floatfix]{revtex4-2}

\usepackage{graphicx}
\usepackage{dcolumn}
\usepackage{bm}
\usepackage{color}
\usepackage[usenames,dvipsnames]{xcolor}
\usepackage{hyperref}
\usepackage{mathtools}
\usepackage{siunitx}
\usepackage{lipsum}
\usepackage[mathlines]{lineno}

\newcommand{\Jc}{J_\mathrm{c}}

\newcommand{\Jdp}{J_\mathrm{dp}}

\newcommand{\Jdc}{J_\mathrm{dc}}

\newcommand{\Jext}{\mathbf{J}}
\newcommand{\JN}{\mathbf{J}_\mathrm{n}}
\newcommand{\JS}{\mathbf{J}_\mathrm{s}}
\newcommand{\Tc}{T_\mathrm{c}}

\newcommand{\Hc}{H_\mathrm{c1}}
\newcommand{\Hct}{H_\mathrm{c2}}

\newcommand{\FF}{\mathrm{FF}}

\begin{document}


\title{ Driven Vortex Motion in Periodic Channel Potential Embedded in Superconducting Films }

\author{A. Al Luhaibi}
\email{aalluhaibi@kfupm.edu.sa}
\affiliation{Physics Department, King Fahd University of Petroleum and Minerals, 31261 Dhahran, Saudi Arabia}
\affiliation{Interdisciplinary Research Center (IRC) Advanced Quantum Computing $,$ KFUPM$,$ Dhahran$,$ Saudi Arabia}

\author{A. Glatz}
\email{glatz@anl.gov}
\affiliation{Materials Science Division, Argonne National Laboratory, 9700 S. Cass Avenue, Argonne, Illinois 60439, USA} 
\affiliation{Department of Physics, Northern Illinois University, DeKalb, Illinois 60115, USA}

\author{Michael Vogl}
\affiliation{Physics Department, King Fahd University of Petroleum and Minerals, 31261 Dhahran, Saudi Arabia}
\affiliation{Interdisciplinary Research Center (IRC) Advanced Quantum Computing $,$ KFUPM$,$ Dhahran$,$ Saudi Arabia}

\author{Hocine Bahlouli}
\affiliation{Physics Department, King Fahd University of Petroleum and Minerals, 31261 Dhahran, Saudi Arabia}
\affiliation{Interdisciplinary Research Center (IRC) Advanced Materials $,$ KFUPM$,$ Dhahran$,$ Saudi Arabia}

\author{J. B. Ketterson}
\email{j-ketterson@northwestern.edu}
\affiliation{Department of Physics and Astronomy, Northwestern University, Evanston, Illinois 60208, USA}

\date{\today}

\begin{abstract}

In this paper, we investigate vortex dynamics in periodically width-modulated superconducting channels by numerically solving the time-dependent Ginzburg–Landau (TDGL) equations. The channels are designed with sinusoidally varying widths characterized by their average width and modulation amplitude. The motion of driven vortices generates an oscillatory electric field with a fundamental frequency, $f_0$, set by the channel periodicity, together with higher-order harmonics. In contrast to the situation of vortex lattice dynamics in a periodic pinning landscape of circular holes considered by \citet{AlLuhaibi:2022}, we show that the harmonic content of the generated electric field can be effectively controlled through the channel geometry. In particular, appropriately engineered channel profiles strongly suppress higher-order harmonics, yielding an electric-field response that closely approaches a pure sinusoidal waveform. These results demonstrate that geometric modulation offers a viable route to engineering low-harmonicity microwave signals in superconducting devices and may enable cryogenic on-chip waveform generators for controlling superconducting qubits. Such devices could provide a useful building block for scalable transmon-based quantum computing architectures.
\end{abstract}


\maketitle



\section{Introduction}\label{sec:Intro}

Superconducting vortices, which contain localized normal-state cores, penetrate a type-II superconductor when the applied magnetic field exceeds the lower critical field, $\Hc$. When an external current density $\Jext$ is applied, each vortex experiences a Lorentz force, $\mathbf{F}_L=\Jext\times\mathbf{H}$, directed perpendicular to both the current and magnetic field~\cite{kopnin}. In the absence of pinning and applied currents, vortices arrange themselves into a hexagonal Abrikosov lattice~\cite{deGennes:Book}. Under an applied current, this lattice can move, producing a finite electric field and hence a dissipative voltage response~\cite{Tinkham:Book}. The motion of vortices can be strongly modified by defects, which act as local pinning centers and impede vortex motion~\cite{Kwok:2016}. In conventional superconducting materials, such defects typically arise from impurities, structural disorder, or other imperfections and are distributed randomly throughout the sample~\cite{Ganguli:2015,Kimmel:2019}. In contrast, artificial pinning landscapes can be deliberately engineered by introducing ordered arrays of defects, for example, through patterned holes (antidots)~\cite{hebard1978critical,raedts2004flux} or by depositing normal-metal dots on the surface of a superconducting film~\cite{hoffmann2000periodic}.

Artificially structured pinning landscapes provide a means of controlling vortex motion and, consequently, the electrical response of superconducting devices. In particular, \citet{AlLuhaibi:2022} investigated vortex dynamics in otherwise uniform superconducting films containing periodic arrays of artificial pinning sites. As driven vortices repeatedly become pinned and depinned while traversing the periodic landscape, their motion produces a periodic voltage response with a well-defined fundamental frequency, often referred to as the hopping frequency, $f_0$. However, the abrupt nature of the pinning and depinning processes introduces strong temporal nonlinearities into the vortex dynamics. Consequently, the resulting voltage waveform is intrinsically non-sinusoidal and contains substantial higher-order harmonic components.

An alternative approach to controlling vortex motion is to replace discrete pinning sites with a continuously varying pinning landscape. Inspired by guiding structures used for fluid transport, superconducting channels with spatially varying geometries can be engineered to guide vortex motion along prescribed paths. Such structures can be fabricated using standard lithographic techniques~\cite{VELEZ20082547,morgan1998asymmetric}. Their geometry can be characterized by parameters such as the channel width, depth, and centerline profile, each of which can be varied spatially along the direction of vortex motion. Continuous geometric modulation therefore provides additional degrees of freedom for controlling vortex dynamics beyond those available in arrays of localized pinning centers.

Despite this potential, vortex dynamics in continuously modulated superconducting channels have received comparatively little attention. To our knowledge, only one experimental study has investigated vortex motion in superconducting channels with periodic constrictions~\cite{Yu:2010}. In the present work, we focus on a particularly simple and experimentally accessible geometry: superconducting channels whose width varies periodically along the direction of vortex motion. This geometry provides a controlled setting in which to investigate how the spatial modulation of the channel influences vortex dynamics and the resulting electric-field waveform.

We first investigate vortex dynamics in uniform superconducting channels for different channel widths and values of the parameter $\epsilon$, which characterizes the strength of the pinning potential arising from the local spatial variation of the critical temperature, $T_c$ (Sec.\ref{sec:UniResp}). We then consider channels with sinusoidally modulated widths and examine how the modulation parameters influence the resulting vortex dynamics and electric-field response (Sec.\ref{sec:SinChanl}). In contrast to systems with discrete localized pinning centers, the continuous and spatially modulated pinning landscape provides a means of modifying the temporal structure of the vortex motion and, consequently, the harmonic content of the generated electric field.

We model the vortex dynamics using the time-dependent Ginzburg--Landau (TDGL) formalism. The numerical implementation and simulation methodology are described in Sec.\ref{sec:TDGL_Sim}, while the superconducting system and channel geometry considered in this work are introduced in Sec.\ref{sec:sample}.  In Sec.~\ref{sec:Analysis}, we develop a systematic theoretical framework for analyzing the vortex dynamics and the resulting electric-field response.

In Sec.~\ref{sec:Resp}, we investigate the response of the system to a dc driving current for both uniform and sinusoidally modulated channels. Sec.~\ref{sec:discussion} we discuss our  main results together with a qualitative physical interpretation. In Sec.~\ref{sec:appl}, we discuss the potential applications of our findings, and finally, Sec.~\ref{sec:conclude} summarizes the main conclusions.

\section{Model \label{sec:model}}

\begin{figure}[htb]
    \centering
    \includegraphics[width=\columnwidth]{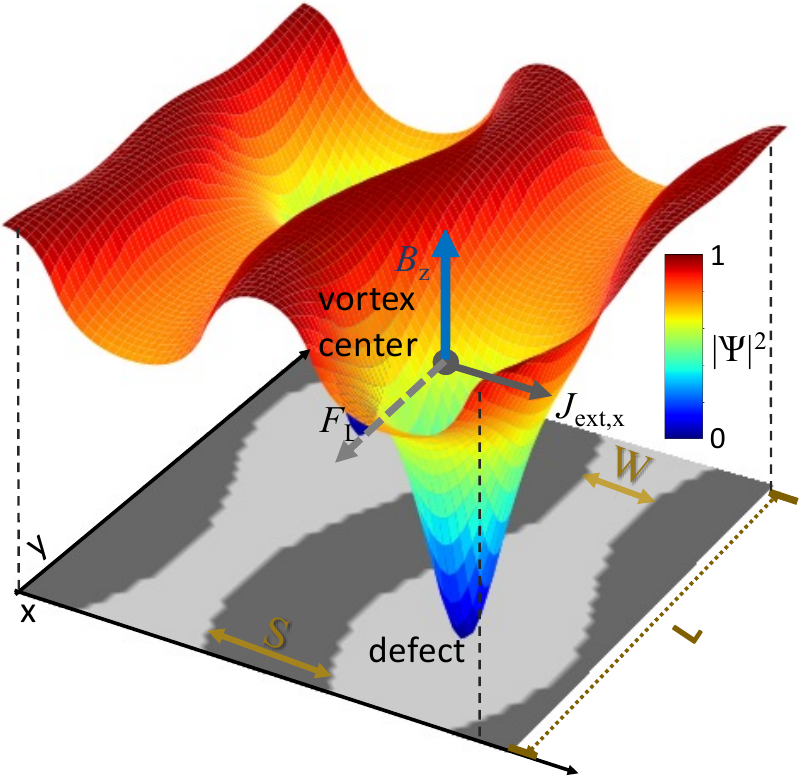}
    \caption{Surface plots of $|\psi|^2$ in the simulated system of size $L\times L$ with two sinusoidal shape channel defects separated by $W$ (separation width). $L$ representing the length of the periodic unit cell. The projection indicates regions with different $\Tc$/$\epsilon$-values in dark gray ($\epsilon=1$) and light gray ($\epsilon=0.75$). At the central position of the vortex, the direction of applied current, magnetic field, and resulting Lorentz force ($F_L$) are indicated. }
    \label{fig:geometry}
\end{figure}

The time-dependent Ginzburg-Landau (TDGL) equations provide a powerful computational tool for describing the dynamics of the superconducting order parameter near the critical temperature and are used to extract the macroscopic behavior of type-II superconductors. In the present work, we employ a recently developed, highly optimized iterative implicit solver for the TDGL equations, introduced by~\citet{Largestable}. An important advantage of this formulation lies in its ability to accommodate arbitrary defect parameters, thereby enabling a systematic study of the influence of channel geometry and pinning strength on vortex dynamics, as presented in Section \ref{sec:TDGL_Sim}. Although a wide variety of channel-edge profiles can, in principle, be considered, we restrict our analysis to a sinusoidally modulated channel due to its geometric simplicity, smoothness, and experimental feasibility (see Fig.~\ref{fig:geometry}). In most of this work, we focus on the sinusoidally modulated channel; its geometry is described in detail in Sec.~\ref{sec:sample}. To better understand the resulting electric field waveforms, we have focused on two parameters: 1) the waveform efficiency defined by the parameter $\chi$, which is defined as the ratio of converted ac power to total power. 2) the harmonicity of the waveform represented by the parameter $\zeta$, which provides the ratio of the fundamental to all the harmonics; for details, see Sec.~\ref{sec:Analysis}.

\subsection{TDGL Simulations}\label{sec:TDGL_Sim} 
The time-dependent Ginzburg-Landau (TDGL) equations for the order parameter, $\psi(r)$, may be written as \cite{Largestable},
\begin{eqnarray}
     \Gamma\left(\partial_t -\imath\frac{2 e}{\hbar}\mu\right)\psi&=& a_0\epsilon(\mathbf{r})\psi - b|\psi|^2\psi+\nonumber\\
     &&\frac{1}{4m}\left(\hbar\nabla+\frac{2e}{\imath c}\mathbf{A}\right)^2\psi\,,\label{equ:GLE1}\\
     \nabla\times\nabla\times \mathbf{A} &=& \frac{4\pi}{c} \left(\JS+\JN\right)\,. \label{equ:GLE2}
\end{eqnarray}
Here $e$ and $m$ are the electron charge and mass, respectively, $c$ is the speed of light in vacuum, $\mu$ and $\mathbf{A}$ are the scalar and vector potentials, $\Gamma$, $a_0$ and $b$ are material dependent parameters, and $\epsilon(\mathbf{r}) \equiv (T_c(\mathbf{r})-T)/T$, where $T$, and $T_c(\mathbf{r})$  are the temperature and the local critical temperature of the superconductor. An important feature of the temperature-dependent parameter $\epsilon(\mathbf{r}) \equiv (T_c(\mathbf{r})-T)/T$
is that it effectively represents the pinning potential within the TDGL model  and vanishes when the temperature reaches the local critical temperature. The quantities $\JN$ and $\JS$ denote the normal and superconducting current densities, respectively, and are given by
\begin{eqnarray}
    \JN &=& -\sigma_n\left(\nabla\mu+\frac{1}{c}\partial_t \mathbf{A}\right)\quad\textrm{and}
    \label{equ:JN}\\
    \JS &=& -\frac{e}{2m}\left(\psi^*\left(\imath \hbar\nabla+\frac{2e}{c}\mathbf{A}\right)\psi +\mathrm{c.c.} \right)\,.
    \label{equ:JS}
\end{eqnarray}

These equations contain two fundamental length scales: the London penetration depth, $\lambda_L(T) = \lambda_{L,0}(1-T/T_c)^{-1/2}$, and the coherence length, $\xi(T) = \xi_0(1-T/T_c)^{-1/2}$; here $\lambda_{L,0}^2=m c^2/(8\pi e^2\psi_0^2)$, with $\psi_0^2 = a_0/b$, and $\xi_0^2=\hbar^2/(4ma_0)$. $\sigma_n$ is the normal conductivity.
In constructing the dimensionless equations, we use the zero-temperature coherence length, $\xi_0$, as the unit of length such that the geometry of inclusions in the superconductor becomes temperature independent. 
The ratio of these two length scales defines the dimensionless Ginzburg-Landau (GL) parameter, $\kappa = \lambda_{L,0}/\xi_0$. Here we use the large $\kappa$-limit, which is suitable for superconducting thin films~\cite{Largestable}.
The relaxation parameter $\Gamma$, which affects the time-evolution, enters via a second dimensionless ratio $u=\Gamma/(\nu_0t_0)$ with $\nu_0$ being the density of states at zero temperature.
With these definitions, the TDGL equations can be written in dimensionless form as follows

\begin{eqnarray}
    u(\partial_t + \imath \mu)\psi &=& \epsilon(\mathbf{r}) \psi - |\psi|^2 \psi + (\nabla - \imath \mathbf{A})^2 \psi\,,
    \label{eq:psi}\\
    \kappa^2 \nabla \times (\nabla \times \mathbf{A}) &=& \JS+\JN\,.
    \label{eq:A}
\end{eqnarray}

Here, the magnetic field is measured in units of the upper critical field $\Hct(0)=\phi_0/(2\pi\xi_0^2)$ ($\phi_0=\pi\hbar c /e$ is the flux quantum). 
The total current density can be written as
\begin{equation}
    \Jext=\JS+\JN=\operatorname{Im} \left[\psi^*(\nabla-\imath \mathbf{A})\psi \right]-(\nabla\mu+\partial_t \mathbf{A})\,,
    \label{eq:Jtot}
\end{equation}
where the unit of the current density is given by $J_0= (e\hbar/m_e\xi_0)\psi_0^2$. In these units, the depairing current density has the value $\Jdp=2/\sqrt{27}\approx 0.385$.

To include pinning forces into our model, we employ the so-called $\delta\Tc$ pinning model~\cite{Kwok:2016}, in which metallic inclusions are modeled through a spatial modulation of the local critical temperature. In Eq.~\eqref{eq:psi}, this is realized through the coefficient of the linear term in the Ginzburg-Landau equation (1). 
In our model, the bulk superconductor is characterized by $\epsilon=1$ whereas within the pinning region we take $\epsilon< 1$; i.e., the local critical temperature of the pinning site is less than the bulk $\Tc$. For values $0<\epsilon<1$ one has weakened superconducting defects, whereas regions with $\epsilon<0$ model normal defects. To model the holes (antidots) discussed here, the deGennes boundary condition $(\nabla - \imath \mathbf{A})\psi\cdot \hat{n} = 0$, is used~\cite{deGennes:Book}, which implies that no supercurrent passes through the defect.

In addition to the supercurrent and normal current, we also apply an external dc current density in $x$-direction.

The externally applied dc current $J_{dc}$ results in a linear contribution to the scalar potential which can then be written as $\mu = -xE+ \Tilde{\mu}(\textbf{r})$, where $E$ is the corresponding electric field and $\Tilde{\mu}(\textbf{r})$ a periodic function used to describe the nonlinear part of the scalar potential. The latter is calculated using the Poisson equation (see Ref.~\cite{Largestable} for details).

It is important to note that the TDGL formalism is valid as long as nonequilibrium excitations are small in number and the system remains in a near-steady state~\cite{kopnin}.
Therefore, we only consider frequencies $f_\mathrm{ext}\ll \Gamma^{-1}=\tau_{\psi}^{-1}$, where $\Gamma=\nu_0\pi\hbar/(8k_B\Tc)$, can be derived from the microscopic theory.
Furthermore, we must keep the applied peak-current density $\Jdc$ well below the depairing current density, $\Jdp$, so that order parameter fluctuations can be neglected.  For a more detailed discussion of non-equilibrium effects, we refer the reader to the book by~\citet{kopnin}.

For the numerical solution of equations~\eqref{eq:psi}, the system is spatially discretized on a regular two-dimensional mesh in space (with a typical grid spacing of $0.3\xi_0$). We perform the time integration using an implicit Crank--Nicolson scheme\cite{Crank_Nicolson_1947} (typical time discretization $0.1t_0$). 
We solve the resulting discretized equations iteratively using the recently developed Jacobi solver  method implemented on graphics processing units (GPUs) for efficient numerical computation.

We impose periodic boundary conditions only along the $x$-direction, corresponding to the direction of the externally applied current. The external current is realized by constraining the spatially averaged total current density along the $y$-direction, following Ref.~\cite{Largestable}. 
The $y$-direction has an open boundary, so the system is designed as a nano-strip geometry.   

\subsection{Designed Pattern}\label{sec:sample}

Most previous studies concentrated on vortex dynamical responses in heterogeneous superconducting systems, focused on  localized pinning sites, which, above the critical current, temporarily trap vortices, causing an abrupt change in the velocity~\cite{Reichhardt:1997,reichhardt:2000,Jiang:2004}. Here, we are interested in a system where the pinning region is not localized but extends across the entire superconducting film along the y-direction, thereby forming what we refer to as a pinning channel. The spatial modulation of the channel width along the transverse $x$-direction induces a continuous variation in the pinning force experienced by the moving vortex, which is directly manifested in the resulting electrical response.

The geometric modulation considered here consists of a sinusoidal variation in the channel width. To synchronize the vortex motion and favor an in-phase dynamical response, every other channel is shifted by half a modulation period relative to its neighboring channels~\cite{AlLuhaibi:2022}, as illustrated in Fig.~.\ref{fig:geometry}. The channel geometry is implemented by assigning different values of the parameter $\epsilon$ to the respective regions, as specified below the figure.

\subsection{Theoretical Analysis}\label{sec:Analysis}

In this work, our primary interest is characterizing the system's oscillatory response, which is best analyzed in the frequency domain using Fourier-transform techniques; hence, we write;
\begin{equation}
    E(t) = \sum_{n = 0}^{n = \infty} A_n e^{\imath2n\pi f_0 }, 
\end{equation}

where $f_0$ denotes the fundamental frequency, and $A_n$ is the amplitude of each harmonic. Note that $A_0$ is the dc component of the response. Although the system is driven by a dc current, the resulting electrical response generally contains both dc and ac components. The dc component contributes to Joule heating and, although relatively small in vortex-based systems, represents a dissipative loss that should ideally be minimized. In contrast, the ac component is associated with generating electromagnetic waves. The ability to control this oscillatory response through vortex dynamics therefore provides a potential mechanism for realizing a tunable electromagnetic-wave generator.

We introduce two dimensionless parameters to characterize the quality and efficiency of the signal generated by the modulated channel. First, to quantify the efficiency of the ac voltage response, we define the conversion efficiency, $\chi$, as the ratio,

\begin{equation}
    \chi = \frac{ \sum_{n = 1}^{n = \infty} |A_n|^2 }{ \sum_{n = 0}^{n = \infty}|A_n|^2 }.
    \label{eq:eff}
\end{equation}

The second parameter, the harmonicity $\zeta$, measures the contribution of the first harmonic $A_1$ relative to all higher-order harmonics $A_n$ with $n>1$, and is defined as 

\begin{equation}
    \zeta = \frac{ |A_1|^2 }{ \sum_{n = 1}^{n = \infty} |A_n|^2 }.
    \label{eq:hrm}
\end{equation}

The interpretation of $\zeta$ is straightforward: as $\zeta$ approaches unity, the contribution of higher-order harmonics becomes negligible and the waveform approaches a purely sinusoidal form. Hence, together, $\chi$ and $\zeta$ provide complementary measures of the efficiency of ac signal generation and the spectral purity of the resulting waveform, respectively.

\section{Responses to dc current \label{sec:Resp}}

In this section, we investigate the response of the system shown in Fig.~\ref{fig:geometry}, under an applied dc current, focusing specifically on the low-current regime ($J<0.1$). 

This section is organized as follows. We first study response of the uniform channel, for which the channel's width is constant (see Sec.~\ref{sec:UniResp}), and then examine the response of a sinusoidally modulated channel in Sec.~\ref{sec:SinChanl}. 

\subsection{Uniform Channels \label{sec:UniResp}}

We first consider the uniform-channel limit of the geometry shown in Fig.~\ref{fig:geometry} for which the channel width is constant, $W(y) = W_0$. Accordingly, the light-gray channel boundaries shown in the figure become straight rather than sinusoidally modulated, thereby constraining the vortices to move along the center of a uniform channel.  In this case, even an infinitesimal dc current is sufficient to set the vortices in motion, indicating the absence of a finite depinning threshold, i.e., $J_c = 0$. Moreover, because the pinning landscape is translationally invariant along the direction of vortex motion, the vortices experience no spatially varying pinning force and consequently move with a constant velocity. The resulting electric-field response is therefore time-independent.

\begin{figure}[htb]
    \centering
    \includegraphics[width=\columnwidth]{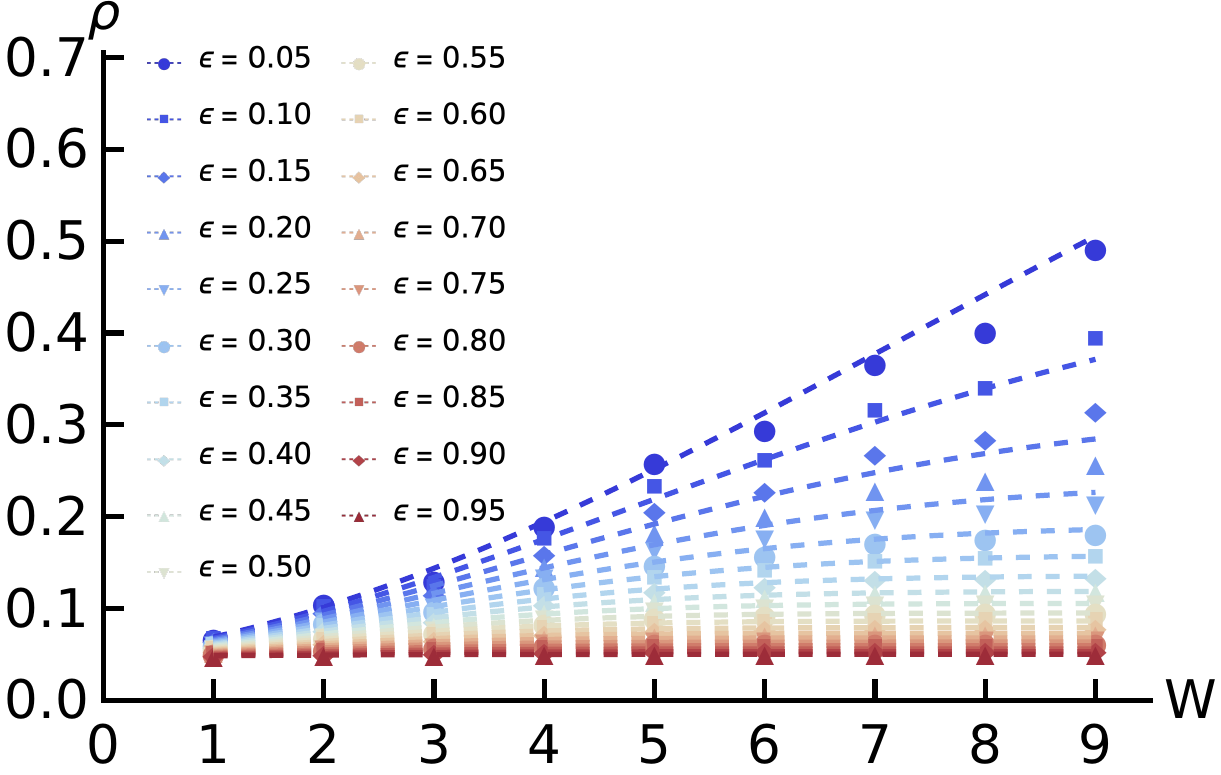}
    \caption{Resistivity as function of the channel's width, $W_0$ for different $\epsilon-$values in a uniform channel.  }
    \label{fig:RvsW}
\end{figure}

Fig.~\ref{fig:RvsW} clearly shows that the resistivity, $\rho$, depends on the channel width and the value of $\epsilon$ . 

These results indicate that spatial variations in the channel parameters, particularly the channel width, $W$, and the local parameter $\epsilon$, directly influence the strength of the resulting electric-field response. 
In particular, it is obvious that in the limit $W\to 0$ the resistivity approaches the free flux flow value for a homogeneous film: $\rho_{\FF}(\epsilon_0)$ (here $\epsilon_0=1$), while for the opposite limit $W_0\to \infty$ it 
approaches $\rho_{\FF}(\epsilon)\propto 1/\epsilon$ ($\epsilon<\epsilon_0$). One also sees that for small $W$, $\rho$ behaves almost linearly with a small higher order contribution and then saturates to $\rho_{\FF}(\epsilon)$ for large $W$.
We therefore introduce the following phenomenological approximation for $\rho$, which captures all these effects:
\begin{eqnarray}
\rho(W,\epsilon_{0},\epsilon)&=&\rho_{\FF}(\epsilon_{0})+\left(\rho_{\FF}(\epsilon)-\rho_{\FF}(\epsilon_{0})\right)\times\label{equ:Rhofit}\\
&&\left[1-e^{-\left[Wf_{1}(\epsilon_{0},\epsilon)+W^{2}f_{2}(\epsilon_{0},\epsilon)\right]}\right]\,,\nonumber
\end{eqnarray}
with two coefficient functions $f_k$. Here $f_2$ captures the deviation from the linear behavior for small $W$.
Since the slope of $\rho$ for small $W$ depends on the ratio of $\rho_{\FF}(\epsilon)$
and $\rho_{\FF}(\varepsilon_{0})$, i.e., how different the asymptontic
limit values are, we use the simple form $f_{k}(\varepsilon_{0},\varepsilon_{c})=a_{k}\rho_{FF}(\varepsilon_{0})/\rho_{\FF}(\varepsilon)$. Using the two parameters $a_{1}$, $a_{2}$ as fitting parameters we can describe the data well with $a_{1}=0.3415$, $a_{2}=0.1366$ (dashed lines in Fig.~\ref{fig:RvsW}).

\subsection{Sinusoidal  Channels \label{sec:SinChanl} }

Next, we investigate channels whose width periodically varies according to sinusoidal modulation.
\begin{equation}
 W = W_0+\alpha \sin(\frac{2\pi}{L}y),  \label{equ:SinWidth}  
\end{equation}
where $W_0$ denotes the average channel width, while $\alpha$ controls the modulation amplitude and, consequently, the maximum and minimum channel widths. The sinusoidal profile is chosen primarily for three reasons: periodicity, simplicity, and smoothness. First, its intrinsic periodicity is naturally compatible with the periodic boundary conditions used in our simulations, thereby minimizing possible artificial boundary effects. Second, the sinusoidal modulation contains a single spatial frequency, providing a simple and well-defined geometry for analyzing the relationship between channel modulation and the resulting dynamical response. Finally, the channel boundaries vary continuously and smoothly, avoiding abrupt geometric variations that could introduce anomalous features into the vortex dynamics. In the following analysis, we use the spacing between adjacent channels as a control parameter because, unlike the local channel width $W$, it remains constant throughout the system. It is worth mentioning that increasing the inter-channel spacing corresponds to decreasing the average channel width $W_0$.

\begin{figure}[htb]
    \centering
    \includegraphics[width=\columnwidth]{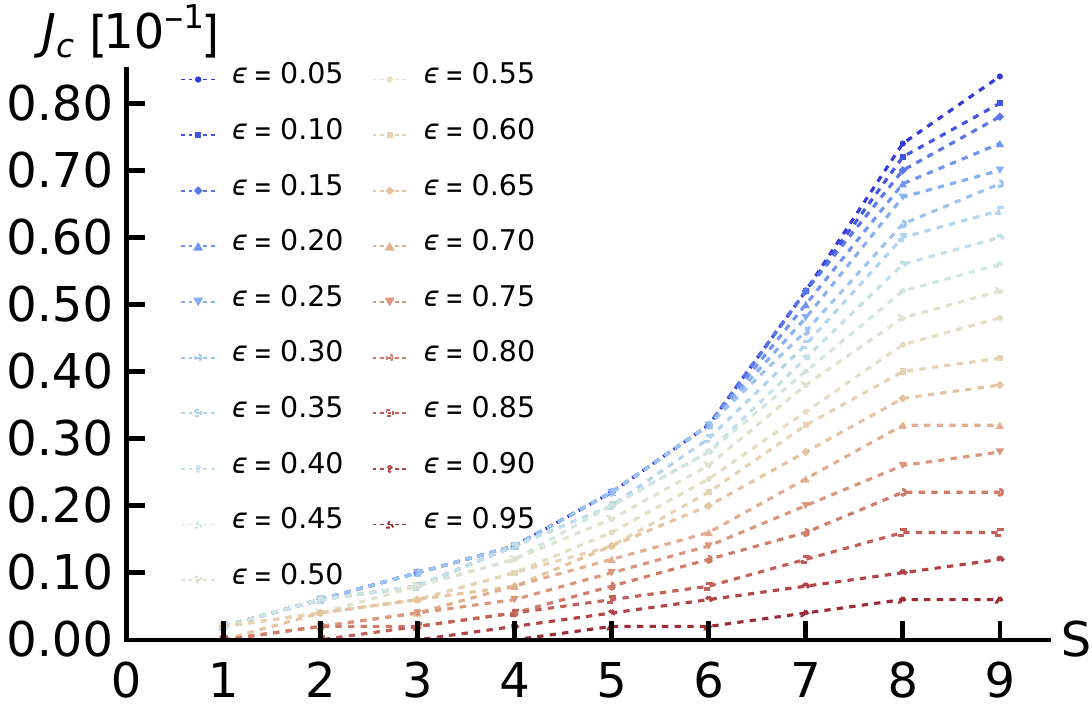}
    \caption{Critical current as a function of the channel spacing for different values of $\epsilon$.  }
    \label{fig:Jc_w_ep}
\end{figure}

The first thing one can investigate is the behavior of the critical current, shown in Fig.~\ref{fig:Jc_w_ep}, for different values of inter-channel spacing $S$ and $\epsilon$, this implies that the potential landscape changes due to variations in the width.

\begin{figure}[htb]
    \centering
    \includegraphics[width=\columnwidth]{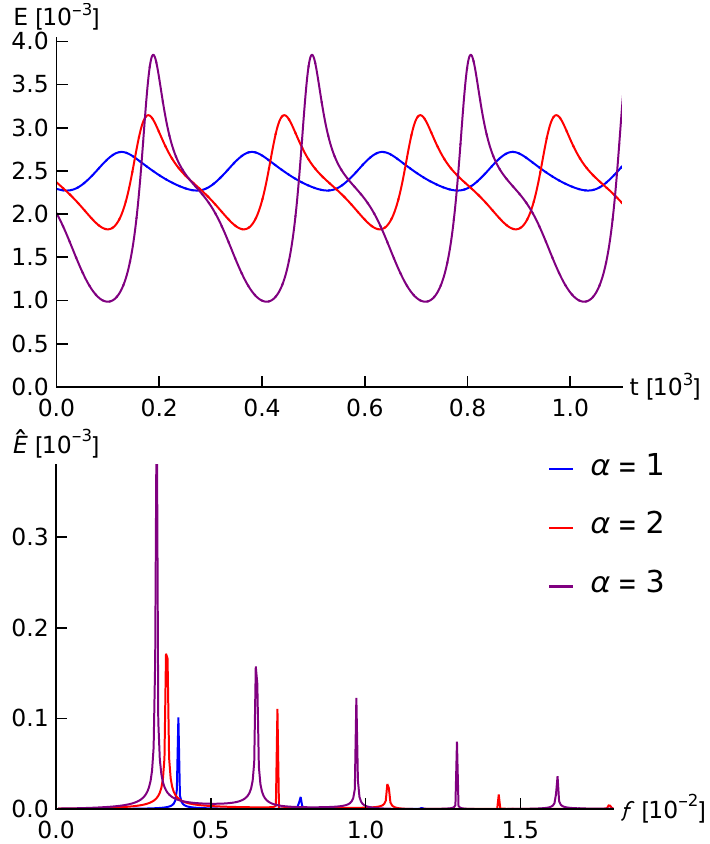}
    \caption{\textit{(top)} Time-dependent electric field curves for $L = 20$, $W = 3$, and $\epsilon = 0.8$ system with $\Jdc=0.04$ for different values of the modulation amplitude, $\alpha $.  \textit{(bottom)} The Fourier transforms of the $E(t)$ curves reveal the generated frequencies. The Fourier analysis also shows higher harmonics.}
    \label{fig:DiffAmp}
\end{figure}

\begin{figure}[htb]
    \centering
    \includegraphics[width=\columnwidth]{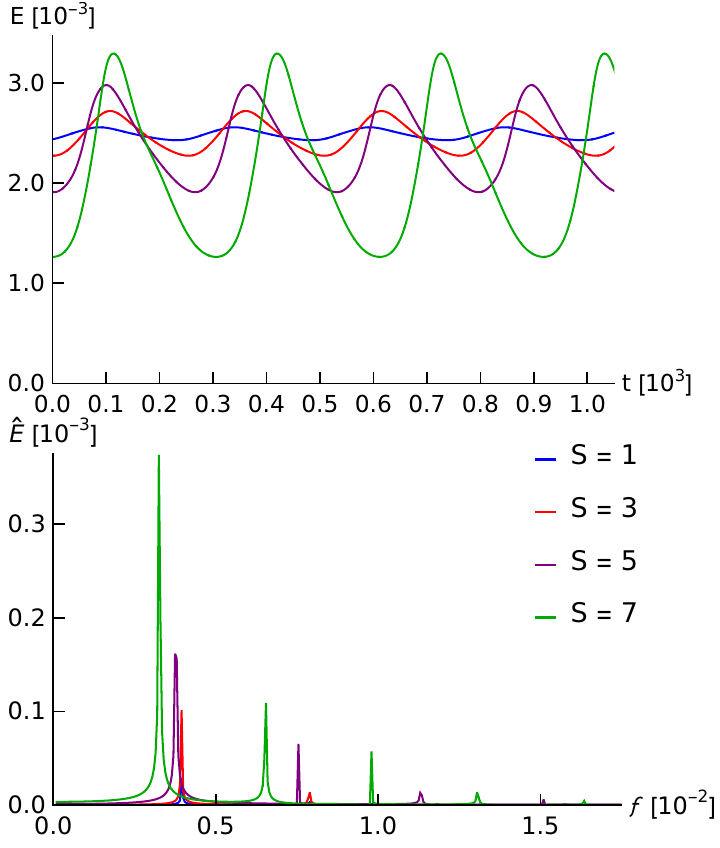}
    \caption{\textit{(top)} Time-dependent electric field curves for $L = 20$, $\alpha = 1$, and $\epsilon = 0.8$ system with $\Jdc=0.04$ for different values of the inter-spacing parameter, S.  \textit{(bottom)} The Fourier transforms of $E(t)$ curves exhibit the generated frequencies. The Fourier analysis also shows higher harmonics.}
    \label{fig:DiffW}
\end{figure}

\begin{figure}[htb]
    \centering
    \includegraphics[width=\columnwidth]{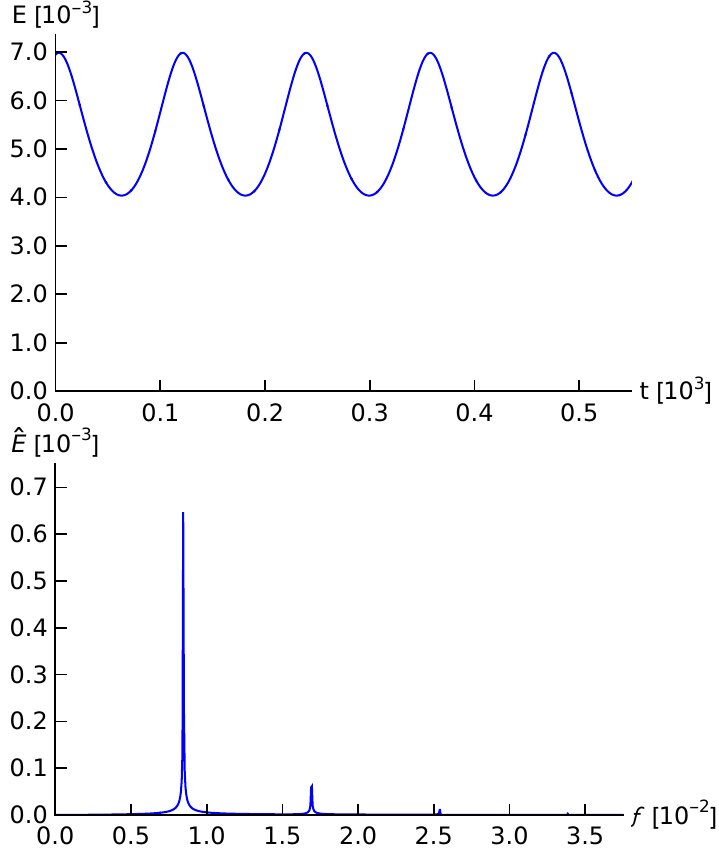}
    \caption{\textit{(top)} Time-dependent electric field curves for the $L = 20$, $S = 1$, and $\epsilon = 0.1$ system with $\Jdc=0.014$.  The response is sinusoidal wave \textit{(bottom)} The Fourier transforms of the $E(t)$ curves expose the generated frequencies. The Fourier analysis shows that the higher harmonics are quite weak in comparison to the fundamental frequency.}
    \label{fig:Sin}
\end{figure}

Varying the parameters of our sinusoidal channel width (see Fig.~\ref{fig:DiffAmp}, and~\ref{fig:DiffW}) changes the waveform dramatically, from being Sinusoidal-like function (see Fig.~\ref{fig:Sin}) to an almost pulse-like function (see Fig.~\ref{fig:DiffAmp}). The results demonstrate that both the modulation amplitude, $\alpha$, and the inter-channel spacing, $S$ (or, equivalently, the average channel width, $W_0$), have a pronounced influence on the resulting waveform, whereas variations in $\epsilon$ produce comparatively modest effects.

We observed that increasing the spacial amplitude reduces the fundamental frequency as shown in Fig.~\ref{fig:DiffAmp}. Similarly, we observe in Fig.~\ref{fig:DiffW} that increasing the inter-channel spacing S reduces the fundamental frequency. Thus, increasing either the modulation amplitude, $\alpha$, or the inter-channel spacing, $S$, produces a qualitatively similar effect on the spectral response, namely, a reduction in the fundamental frequency.

The existence of a sinusoidal waveform always persists when we increase the applied current even in the case of localized pinning, because a fast moving vortex is only weakly influenced by the pinning potential landscape. A simple example that supports this observation is the tilted-washboard potential (see reference \cite{Tinkham:Book} for details). 

\begin{figure}[htb]
    \centering
    \includegraphics[width=\columnwidth]{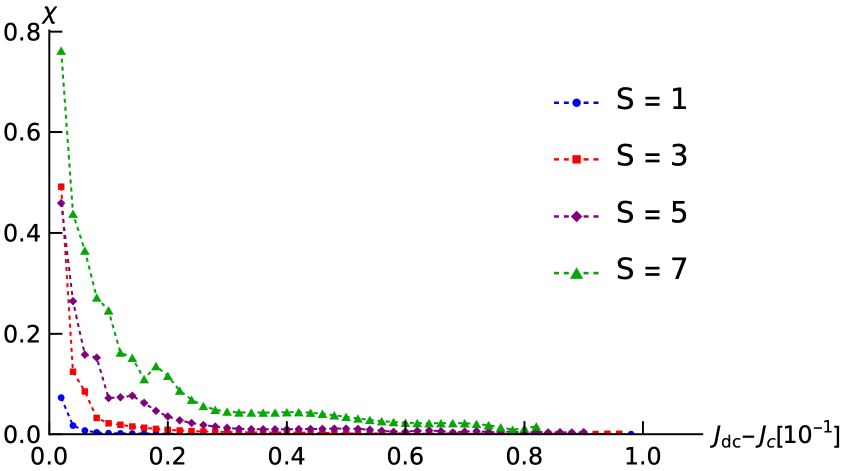}
    \caption{The efficiency, $\eta$, vs the applied current above the critical current, $J-\Jc$, for different values of the inter-spacing parameter, S. The system size $L$ = 20, and the channel has  $\epsilon$ = 0.8.    }
    \label{fig:effvsF}
\end{figure}

\begin{figure}[htb]
    \centering
    \includegraphics[width=\columnwidth]{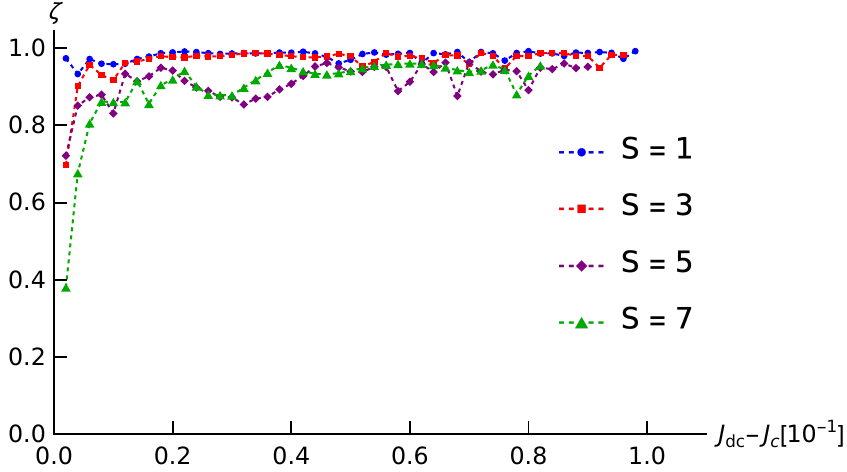}
    \caption{The harmonicity, $\zeta$, vs the applied current above the critical current, $J-\Jc$, for different values of the inter-spacing parameter, S. The system size $L$ = 20, and the channel has  $\epsilon$ = 0.8.  }
    \label{fig:HrmvsF}
\end{figure}

However, increasing the applied current always affects the efficiency for generating more sinusoidal waves while the remaining energy is lost  as heat to the system, which weakens its potential for sinusoidal wave generator applications. An important target of this work is to generate sinusoidal waveforms at the highest efficiency possible. As can be seen in Fig.~\ref{fig:effvsF} and ~\ref{fig:HrmvsF}, the relation between efficiency (Eq.~\eqref{eq:eff}) and harmonicity (Eq.~\eqref{eq:hrm}), is as expected. However, $\epsilon$ has a pronounced effect on the signal amplitude but only a relatively weak influence on the efficiency and harmonicity. In general, simultaneously achieving high harmonicity and high conversion efficiency remains challenging, reflecting a trade-off between spectral purity and power conversion. Nevertheless, in Fig.~\ref{fig:Sin}  one can identify several parameter selections that yield a nearly sinusoidal waveform with reasonable conversion efficiency.

\section{Discussion}\label{sec:discussion}

Fig.~\ref{fig:RvsW} reveals that, depending on the value of $\epsilon$ the resistivity increases monotonically with increasing channel width, whereas this dependence becomes progressively weaker and eventually the resistivity saturates as $\epsilon \rightarrow 1$. For most values of $\epsilon$ the curves exhibit a plateau at sufficiently large channel widths, corresponding to the homogeneous limit in which finite-width effects become negligible. In this regime, the resistivity approaches that of a homogeneous film with the corresponding value of $\epsilon$. This behavior is expected to survive for small values of $\epsilon$; however, the channel widths required to reach the asymptotic homogeneous limit exceed the range explored in the present simulations.  

The free energy density of a uniform channel, $W(y)=W_0$, is translationally invariant along the $y$-direction and therefore produces no pinning-force along y-direction, resulting in a constant electric field response. However, as shown in Fig.~\ref{fig:RvsW}, different channel widths produce different resistivities because geometric confinement modifies the effective vortex viscosity $\eta$. In particular, narrower channels generally lead to larger effective viscosities, since flux-flow resistivity is inversely proportional to vortex viscosity, $\rho \propto \frac{1}{\eta}$, as implied by the Bardeen–Stephen model,~\cite{Tinkham:Book,Bardeen:1965}. 

Simulations of both uniform and sinusoidal channels indicate that the oscillatory electric-field response for sinusoidal channels arises from two distinct mechanisms: the local channel width, as illustrated in Fig.~\ref{fig:RvsW}, and the rate at which the width varies along the channel. The latter determines the pinning force and is controlled by the parameter $\alpha$ in equation~\eqref{equ:SinWidth}. 

A particularly simple regime is obtained when $\frac{\alpha}{W_0} \ll 1$, 
corresponding to a large average channel width or, equivalently, a small relative modulation of the channel width. In this limit, the pinning force along the $y$-direction becomes negligible, and the dynamical response is governed primarily by the resistivity, as described by equation~\eqref{equ:Rhofit}. For this simple situation, see Eq.~\eqref{equ:SinWidth}, the resistivity becomes 
\begin{equation}
   \rho(W) \approx \rho(W_0) +  \beta\sin(\frac{2\pi}{L}y), 
   \label{equ:R_simple}
\end{equation}

where $\beta = \left(\rho_{\FF}(\epsilon)-\rho_{\FF}(\epsilon_{0})\right)f_{1}(\epsilon_{0},\epsilon)\alpha$. 
\\
\\
The electric-field response can then be written as $\textbf{E} = (\rho(W_0) +  \beta\sin(\frac{2\pi}{L}y) )\Jext$. Assuming that the vortex moves with approximately constant velocity through the channel, the resulting electric field exhibits an oscillatory response arising solely from the spatial modulation of the resistivity. Thus, even in the absence of an appreciable longitudinal pinning force, the periodic channel geometry can generate an oscillatory electrical signal.

Thus, under a constant applied current, the temporal modulation of the electric-field response is governed primarily by the spatial variation of the effective resistivity. Consequently, in this regime, the electric-field waveform is expected to closely reflect the spatial modulation of the channel width, $W(y)$.   

It is important to emphasize that this simplified phenomenological model is valid only in the weak-modulation regime $\frac{\alpha}{W_0} \ll 1$. A more comprehensive study is required to fully investigate vortex dynamics in strongly modulated channels and to develop a better theoretical description of the resulting electrical response.

\section{Application}\label{sec:appl}

The simulations presented here show that applying a constant current density produces an oscillating electric field. Waveforms and frequencies depend crucially on the channel's geometry. In particular, tailoring the channel shape can strongly suppress higher harmonics, producing a nearly sinusoidal signal, while other channel geometries produce different pulse shapes. These results suggest a possible application in superconducting-qubit control that uses microwave signals~\cite{krantz:2019}.

Further development of this approach could enable a cryogenic microwave waveform generator. Superconducting qubits operate at millikelvin temperatures but are typically controlled by room-temperature microwave electronics~\cite{krantz:2019}. Delivering these signals requires cables between the two temperature stages; as the number of qubits grows, the cables become harder to accommodate and increase the low-temperature heat load. This challenge has motivated approaches that generate control signals closer to the qubits, including proposals to use trains of discrete single-flux-quantum pulses~\cite{McDermott:2014}. The geometry-controlled signals predicted in this work suggest another possible approach. Whether they can be coupled to a qubit and delivered with the timing required for qubit operations remains to be investigated.


\section{Conclusion}\label{sec:conclude}

Using the time-dependent Ginzburg--Landau (TDGL) model, we have demonstrated that periodically modulating the pinning landscape through variations in channel width gives rise to an oscillatory electrical response. The resulting waveform can be controlled through the channel parameters, particularly the channel's spatial profile $W(y)$, describing variations in the channel width, and the local parameter $\epsilon$, which represents the strength of the pinning potential. We first investigated uniform-width channels and established the dependence of the resistivity on both channel width $W(y)$ and pinning strength $\epsilon$, as shown in Fig.~\ref{fig:RvsW}. These results provide a basis for the interpretation of the more complex electrical responses obtained from the TDGL simulations in the case of width-modulated channels. We then developed a simplified phenomenological model for the weak-modulation regime, in which the average channel width is much larger than the modulation amplitude. In this limit, the longitudinal pinning force becomes negligible, and the electric-field response is governed, primarily, by the spatial variation of the effective vortex viscosity. Consequently, the electric-field waveform closely follows the spatial profile of the channel width, $W(y)$.

These findings demonstrate the potential of geometrically modulated vortex channels as integrated cryogenic waveform generators. By enabling on-chip generation of tailored microwave signals, such devices offer a promising route toward scalable superconducting quantum technologies, where waveform generation and qubit-control circuitry can be cohesively integrated on the same quantum processor.
Finally, further theoretical and experimental investigations are required to optimize this approach and establish the feasibility of generating efficient, precise, and controllable waveforms.

\section{Acknowledgments}
A. Al Luhaibi at KFUPM would like to acknowledge the support received under the University Funded Grant EC241024. 
H. B. acknowledge the support of physics department and the IRC-Advanced Materials during the progress of this work.
The research by AG was sponsored by the Army Research Office and was accomplished under Grant Number W911NF-24-1-0145. 
The views and conclusions contained in this document are those of the authors and should not be interpreted as representing 
the official policies, either expressed or implied, of the Army Research Office or the U.S. Government. The U.S. Government 
is authorized to reproduce and distribute reprints for Government purposes notwithstanding any copyright notation herein.


\bibliography{V_Channel}

\end{document}